\documentclass[longauth]{aa}
\usepackage{amssymb}
\usepackage{amsmath}
\usepackage{xspace}
\usepackage{graphicx}
\usepackage{enumerate}   
\usepackage{color}
\usepackage{txfonts}
\usepackage{tablefootnote}
\usepackage{physics}
\usepackage[switch]{lineno}

\def\es{1ES~0229+200}

\usepackage{natbib}

\def\gr{$\gamma$-ray}
\def\GR{$\gamma$ ray}

\def\lat{\textit{Fermi}/LAT}

\usepackage{color}

\begin{document}

\title{A lower bound on intergalactic magnetic fields from time variability of 1ES~0229+200 from MAGIC and Fermi/LAT observations}

\author{
V.~A.~Acciari\inst{1} \and
I.~Agudo\inst{15} \and
T.~Aniello\inst{2} \and
S.~Ansoldi\inst{3,43} \and
L.~A.~Antonelli\inst{2} \and
A.~Arbet Engels\inst{4} \and
M.~Artero\inst{5} \and
K.~Asano\inst{6} \and
D.~Baack\inst{7} \and
A.~Babi\'c\inst{8}
% ...
}
\institute { Instituto de Astrof\'isica de Canarias and Dpto. de  Astrof\'isica, Universidad de La Laguna, E-38200, La Laguna, Tenerife, Spain
\and National Institute for Astrophysics (INAF), I-00136 Rome, Italy
\and Universit\`a di Udine and INFN Trieste, I-33100 Udine, Italy
\and Max-Planck-Institut f\"ur Physik, D-80805 M\"unchen, Germany
\and Institut de F\'isica d'Altes Energies (IFAE), The Barcelona Institute of Science and Technology (BIST), E-08193 Bellaterra (Barcelona), Spain
\and Japanese MAGIC Group: Institute for Cosmic Ray Research (ICRR), The University of Tokyo, Kashiwa, 277-8582 Chiba, Japan
\and Technische Universit\"at Dortmund, D-44221 Dortmund, Germany
\and Croatian MAGIC Group: University of Zagreb, Faculty of Electrical Engineering and Computing (FER), 10000 Zagreb, Croatia
\and IPARCOS Institute and EMFTEL Department, Universidad Complutense de Madrid, E-28040 Madrid, Spain
\and Centro Brasileiro de Pesquisas F\'isicas (CBPF), 22290-180 URCA, Rio de Janeiro (RJ), Brazil
\and Universit\`a di Padova and INFN, I-35131 Padova, Italy
\and University of Lodz, Faculty of Physics and Applied Informatics, Department of Astrophysics, 90-236 Lodz, Poland
\and Deutsches Elektronen-Synchrotron (DESY), D-15738 Zeuthen, Germany
\and ETH Z\"urich, CH-8093 Z\"urich, Switzerland
\and Instituto de Astrof\'isica de Andaluc\'ia-CSIC, Glorieta de la Astronom\'ia s/n, 18008, Granada, Spain
\and Universit\`a di Pisa and INFN Pisa, I-56126 Pisa, Italy
\and Universitat de Barcelona, ICCUB, IEEC-UB, E-08028 Barcelona, Spain
\and Armenian MAGIC Group: A. Alikhanyan National Science Laboratory, 0036 Yerevan, Armenia
\and Centro de Investigaciones Energ\'eticas, Medioambientales y Tecnol\'ogicas, E-28040 Madrid, Spain
\and Department for Physics and Technology, University of Bergen, Norway
\and INFN MAGIC Group: INFN Sezione di Catania and Dipartimento di Fisica e Astronomia, University of Catania, I-95123 Catania, Italy
\and INFN MAGIC Group: INFN Sezione di Torino and Universit\`a degli Studi di Torino, I-10125 Torino, Italy
\and INFN MAGIC Group: INFN Sezione di Bari and Dipartimento Interateneo di Fisica dell'Universit\`a e del Politecnico di Bari, I-70125 Bari, Italy
\and Croatian MAGIC Group: University of Rijeka, Department of Physics, 51000 Rijeka, Croatia
\and Universit\"at W\"urzburg, D-97074 W\"urzburg, Germany
\and Finnish MAGIC Group: Finnish Centre for Astronomy with ESO, University of Turku, FI-20014 Turku, Finland
\and Departament de F\'isica, and CERES-IEEC, Universitat Aut\`onoma de Barcelona, E-08193 Bellaterra, Spain
\and Japanese MAGIC Group: Physics Program, Graduate School of Advanced Science and Engineering, Hiroshima University, 739-8526 Hiroshima, Japan
\and Armenian MAGIC Group: ICRANet-Armenia at NAS RA, 0019 Yerevan, Armenia
\and Croatian MAGIC Group: University of Split, Faculty of Electrical Engineering, Mechanical Engineering and Naval Architecture (FESB), 21000 Split, Croatia
\and Croatian MAGIC Group: Josip Juraj Strossmayer University of Osijek, Department of Physics, 31000 Osijek, Croatia
\and Japanese MAGIC Group: Department of Physics, Kyoto University, 606-8502 Kyoto, Japan
\and Japanese MAGIC Group: Department of Physics, Tokai University, Hiratsuka, 259-1292 Kanagawa, Japan
\and Universit\`a di Siena and INFN Pisa, I-53100 Siena, Italy
\and Saha Institute of Nuclear Physics, HBNI, 1/AF Bidhannagar, Salt Lake, Sector-1, Kolkata 700064, India
\and Inst. for Nucl. Research and Nucl. Energy, Bulgarian Academy of Sciences, BG-1784 Sofia, Bulgaria
\and Japanese MAGIC Group: Department of Physics, Yamagata University, Yamagata 990-8560, Japan
\and Finnish MAGIC Group: Astronomy Research Unit, University of Oulu, FI-90014 Oulu, Finland
\and Croatian MAGIC Group: Ru\dj{}er Bo\v{s}kovi\'c Institute, 10000 Zagreb, Croatia
\and INFN MAGIC Group: INFN Sezione di Perugia, I-06123 Perugia, Italy
\and INFN MAGIC Group: INFN Roma Tor Vergata, I-00133 Roma, Italy
\and Japanese MAGIC Group: Department of Physics, Konan University, Kobe, Hyogo 658-8501, Japan
\and also at International Center for Relativistic Astrophysics (ICRA), Rome, Italy
% ...
}

\titlerunning{IGMF lower limit from time variability of 1ES~0229+200}

\abstract
{
Extended and delayed emission around distant TeV sources induced by the effects of propagation of \GR s through the intergalactic medium can be used for the measurement of the intergalactic magnetic field (IGMF). 
}
{
We search for delayed GeV emission from the hard-spectrum TeV \gr\ emitting blazar \es, with the goal to detect or constrain the IGMF-dependent secondary flux generated during the propagation of TeV \GR s through the intergalactic medium. 
}
{ 
We analyze the most recent MAGIC observations over a 5 year time span, and complement them  with historic data of the H.E.S.S. and VERITAS telescopes along with a 12-year long exposure of the \lat\  telescope. We use them to trace source evolution in the GeV-TeV band over one-and-a-half decade in time. We use Monte Carlo simulations to predict the delayed secondary \gr\ flux, modulated by the source variability, as revealed by TeV-band observations. We then compare these predictions for various assumed IGMF strengths to all available measurements of the \gr\ flux evolution.
}
{
We find that the source flux in the energy range above 200~GeV experiences  variations around its average on the 14~years time span of observations. No evidence for the flux variability is found in $\mathrm{1-100}$~GeV energy range accessible to \lat. Non-detection of variability due to delayed emission from electromagnetic cascade developing in the intergalactic medium imposes a lower bound of $\mathrm{B>1.8\times 10^{-17}}$~G for long correlation length IGMF and $\mathrm{B>10^{-14}}$~G for an IGMF of the cosmological origin. Though weaker than the one previously derived from the analysis of \lat\  data, this bound is more robust, being based on a conservative intrinsic source spectrum estimate and accounting for the details of source variability in the TeV energy band. We discuss implications of this bound for cosmological magnetic fields which might explain the baryon asymmetry of the Universe.
}
{}

\maketitle
\section{Introduction}

TeV \GR s propagating from distant extragalactic sources suffer from attenuation due to the pair production in interactions with the extragalactic background light (EBL). The pair production effect leads to the deposition of electron-positron pairs in the intergalactic medium. These pairs lose their energy via inverse Compton scattering of cosmic microwave background (CMB) photons and in this way produce a secondary \gr\ flux which could be detected as extended and time-delayed \gr\ ``glow'' around extragalactic TeV \gr\ sources \citep{aharonian_coppi}. Properties of this secondary \gr\ flux are sensitive to weak magnetic fields present in the intergalactic medium \citep{plaga95,neronov07,NeronovSemikoz09,Neronov_Elyiv_10}. This provides us an opportunity to measure such weak magnetic fields using \gr\ observation techniques.

Non-detection of extended or delayed GeV band emission from several extragalactic TeV sources has been previously used to derive a lower bound on the intergalactic magnetic field (IGMF) strength. Indeed, the lower bound on IGMF represents the minimum magnetic field strength needed to suppress the cascade emission in the GeV domain making it negligible with respect to the source emission. The bound on long-correlation-length magnetic field, $\mathrm{B>10^{-16}}$~G~\citep{Neronov10,Tavecchio11} has been derived under the assumption that the present day measurements of the TeV \gr\ flux from selected blazars (a class of active galactic nuclei -- AGNs -- with the jet pointing towards the observer) are representative for the entire AGN duty cycle period of the order of 10~Myr. A weaker bound of $\mathrm{B>10^{-17}}$~G was derived after relaxing this assumption and assuming instead that the selected TeV sources were active only during the historical TeV \gr\ observation time span~\citep{dermer11,Taylor11,Vovk12}. An update of this approach yielding a lower bound $\mathrm{B>3\times 10^{-16}}$~G on long-correlation-length magnetic fields has been reported by \citet{fermi_limit}, based on 7.5 years of the \lat\  telescope data~\citep{FermiLAT}. This updated analysis was also based on the unverified assumption of stability of the TeV band flux on a decade time span.  Limits on extended emission around bright blazars Mrk~421 and Mrk~501 were also derived from the MAGIC telescope data \citep{MAGIC_extended} at energies above 300~GeV, excluding IGMF strengths in the range $\mathrm{4\times10^{-15}~\mathrm{G} - 1.3\times10^{-14}~\mathrm{G}}$, under a strong assumption of existence of persistent intrinsic source flux above 20 TeV. Several other blazars were observed by the H.E.S.S. telescopes~\citep{HESS_IGMF}, yielding similar results -- IGMF strengths in the range $\mathrm{(0.3-3)\times10^{-15}}$~G were excluded, under similar assumptions regarding (unmeasured) intrinsic source fluxes above 20 TeV.

In what follows we report a significant improvement upon these previous results. We use the data of systematic long-term GeV-TeV band monitoring data on a specific source providing the most stringent IGMF lower limits to date, \es, owing to its distance~\citep[redshift of $\mathrm{z\approx 0.14}$,][]{1ES0229_redshift}, hard GeV-TeV spectrum and apparent absence of the high-energy cut-off~\citep{HESS_1ES0229}. The data collected over more than a decade time span allow to substantially relax assumptions about the intrinsic source flux properties over that period. The TeV band monitoring data allow us to make precise prediction of the cascade flux timing properties at lower energies, i.e. those accessible to the \lat\  telescope. The combination of the imaging atmospheric telescope and \lat\  measurements allows us to test the model prediction against the data. We show that non-detection of delayed cascade emission in \lat\  data yields a robust lower bound on IGMF strength and correlation length, free of the assumptions above.

This paper is organized as follows. In Section~\ref{sect::data_analysis} we describe the reduction of the MAGIC and \lat\  data, followed by the numerical modelling of the time-delayed cascade emission. In Section~\ref{sect::analysis_results} we assess the \es\ variability and apply the developed model to evaluate the corresponding bounds on IGMF strength. Finally in Section~\ref{sect::discussion} we discuss the implications of this measurement for the IGMF origin and outline prospects for its further improvements with future observations.

\section{Data analysis}
\label{sect::data_analysis}

\subsection{MAGIC data analysis}

MAGIC is a stereoscopic system of two 17 m diameter Imaging Atmospheric Cherenkov Telescopes (IACT). It is located at 2200 m a.s.l. in the observatory of the Roque de los Muchachos, on the Canary island of La Palma, Spain. It can register \GR s from about 50~GeV to more than 50~TeV. For low-zenith angle observations, the sensitivity of the telescopes for point sources reaches 0.7\% of the Crab Nebula flux above 220 GeV in 50~hr of exposure~\citep{MAGIC_performance}.

MAGIC observations of the \es\, were conducted during the period September 2013 -- December 2017\footnote{The spectral energy distribution  obtained processing the data sample from 2013 to January 2017 has been already published by MAGIC coll. \citep{Acciari2020}. Here we added the data taken in 2017 showing also the overall lightcurve.}. The data set used in our analysis has 145.5~hr of accumulated exposure with zenith angle below 50 degrees, where the MAGIC energy threshold is lower.

The data were analysed using the standard MAGIC software MARS~\citep{zanin_mars_2013}. Standard event cuts are used to improve the signal-to-background ratio as described in~\citet{MAGIC_performance}.
We use standard MAGIC angular cuts to select the source events in the field of view as for time delays shorter than ~$\mathrm{\sim 10}$~years, discussed below, no significant cascade emission is expected to extend beyond the MAGIC point spread function (PSF).

\subsection{\lat\  data analysis}
\label{sect:lat_analysis}

Our analysis of \lat\  data is based on the {\tt P8R3 SOURCE} type \gr\ event selection in the energy range 100~MeV to 200~GeV collected between August 2008 and September 2020. We have processed the data using the Fermitools package and FermiPy framework\footnote{\url{https://fermi.gsfc.nasa.gov/ssc/data/analysis/}} v0.17.3~\citep{fermipy} as described in FermiPy documentation\footnote{\url{https://fermipy.readthedocs.io/}}. 

To extract the source spectrum we have considered the events in the $\mathrm{25^\circ}$-wide region-of-interest around the source position, collected under the zenith angles below $\mathrm{90^\circ}$. We have accounted for the galactic ({\tt gll\_iem\_v07.fits}) and extragalactic ({\tt iso\_P8R3\_SOURCE\_V2\_v1.txt}) diffuse backgrounds and included other sources from the \lat\ fourth source catalogue~\citep[4FGL,][]{4FGL}. The shapes of the sources' spectra were taken from the 4FGL~catalogue with their normalisation left free. We constructed likelihood components separately for each of the four \lat\  PSF classes (event types 4, 8, 16 and 32) and used them jointly in the fit, accounting for the instrument's energy dispersion. The spectra and lightcurves were extracted using two complementary techniques: the binned likelihood analysis and the aperture photometry (for the latter a smaller ~$\mathrm{5^\circ}$ region and ``source'' event type 3 selection was used). The consistency of the results was verified via the crosscheck of the spectra and lightcurves obtained with these two techniques.

We restrict the aperture photometry analysis to the energy range 1-200~GeV (the highest energy photon from the source direction has the energy $\mathrm{E\simeq 180}$~GeV). All over this energy range, the diffuse background event count becomes comparable or lower than that of the source only in the narrow angular bin range $\mathrm{0<\theta<0.25^\circ}$. Taking this into account, we choose the region of extraction of the source signal $\mathrm{\theta<0.25^\circ}$ in the aperture photometry method of spectral extraction. Given that the 95\% containment radius of the \lat\  PSF\footnote{http://www.slac.stanford.edu/exp/glast/groups/\newline /canda/lat\_Performance.htm} above 2~GeV is $\mathrm{\approx 1.5^\circ}$, we use the {\it gtexposure} tool with parameter {\tt apcorr=y} which assures that the source flux estimate from small region of $\mathrm{\theta<0.25^\circ}$ is properly corrected for the source flux fraction contained in this circle.

\subsection{Numerical modelling}
\label{sect:numeric_modelling}

Propagation of \GR s through the intergalactic medium leads to the development of electromagnetic cascades initiated by pair production of the highest energy \GR s on the far-infrared photons of the extragalactic background light. The cascade emission appears as a delayed \gr\ emission following an intrinsic flare of a \gr\ source. To model the delayed emission signal, we have used two fully 3D Monte Carlo codes which trace the development of the cascade in the intergalactic medium: the \texttt{CRPropa} v3.1.7 code \citep{CRPropa} and the \texttt{CRbeam} code developed in~\citet{Berezinsky:2016feh}, also tested via comparison with the alternative cascade simulations~\citep{Taylor11,Kalashev:2014xna,Kachelriess:2011bi}. Detailed comparison of these codes has been performed in~\citet{CRbeam}, where it was shown that simulations with CRPropa and CRbeam agree with each other with an accuracy of the order of 10\% for relatively nearby sources with $z < 0.3$; for the modelling below we have fixed the issue with the inverse Compton interaction rate in CRPropa, identified there.
We use these codes to calculate the cascade signal
$G(E_{\gamma, 0},E_\gamma,t,\tau, B,\lambda_B)$
at the  energy $\mathrm{E_\gamma}$ produced by propagation of primary \GR s of energy $\mathrm{E_{\gamma,0}}$, injected instantaneously by the source and arriving at the time $\mathrm{t}$. The codes give the estimate of the cascade signal as a function of the time delay $\mathrm{\tau}$. The cascade "Green's function" $\mathrm{G}$ depends on the strength $\mathrm{B}$ and correlation length $\mathrm{\lambda_B}$ of IGMF and, in principle, also on the spatial-domain power spectrum of the Fourier expansion of $\mathrm{B}$. Throughout the paper we assume all of the IGMF power is concentrated in a single mode at $\mathrm{k=1/\lambda_B}$ scale.

The cascade flux at an energy $\mathrm{E_\gamma}$ in the observer's frame, based on the known flux variability pattern $\mathrm{F_{s}(E_{\gamma,0},t)}$ can be then expressed as follows:
\begin{equation}
    F_{c}(E_\gamma,t)=
    \int_0^\infty\int_{E_\gamma}^\infty G(E_{\gamma,0},E_\gamma,t-\tau,\tau) F_s(E_{\gamma,0},t-\tau) \dd{E_{\gamma,0}}  \dd{\tau}
\end{equation}

In our calculations we parametrized the intrinsic source light curve (namely the variability pattern) $\mathrm{F_s}$ as 14 steps (bins) in the time interval MJD~53000 -- MJD~59000, covering all of the IACTs and \lat\  observations. The step size is chosen so that to suppress the appearance of the bins with no observations counterparts while still matching the observed source variability (see Sect.~\ref{sect::analysis_spectrum_lightcurve} below). Combined contribution of these time bins (as described by the ``Green's function'' above) was used to estimate both primary and delayed fluxes in each time and energy bin.

As available data do not reveal any variability of the \es\ spectral shape in the $\mathrm{\gtrsim 100}$~GeV energy range, when converting $\mathrm{F_s}$ to primary/secondary flux estimates, we have made a simplifying assumption that the intrinsic source spectrum follows the power law shape with an exponential cut-off
\begin{equation}
    \frac{dN}{dE} = A \left(\frac{E}{E_0}\right)^{-\Gamma} \exp\left(-\frac{E}{E_{cut}}\right)
\end{equation}
and varies only in normalisation (see Sect.~\ref{sect:minimal_cascade} for the details on the chosen spectral shape parameters).

To simulate the secondary emission for \es\ with CRPropa we have assumed the source distance of $\mathrm{D_s = 580}$~Mpc (corresponding to the sources' redshift of $\mathrm{z\approx0.14}$) and recorded all photons injected within a $\mathrm{10^\circ}$ cone, arriving to the sphere of the same radius $\mathrm{r=D_s}$ and centered on the source. CMB and the far infrared backgrounds~\cite[model from][]{Franceschini08} served as the target fields for both \gr\ photon absorption and Inverse Compton scattering of the created secondary $\mathrm{e^+ / e^-}$ pairs. All particles were traced with the required relative integration step accuracy of $\mathrm{\epsilon = 10^{-12}}$ and a minimal step size of $\mathrm{\Delta d = 10^{14}}$~cm, which are sufficient to reproduce time delays with an accuracy better than 1~day.

\section{Analysis results}
\label{sect::analysis_results}

\subsection{Spectrum and light curve}
\label{sect::analysis_spectrum_lightcurve}

\begin{figure}
    \includegraphics[width=\linewidth]{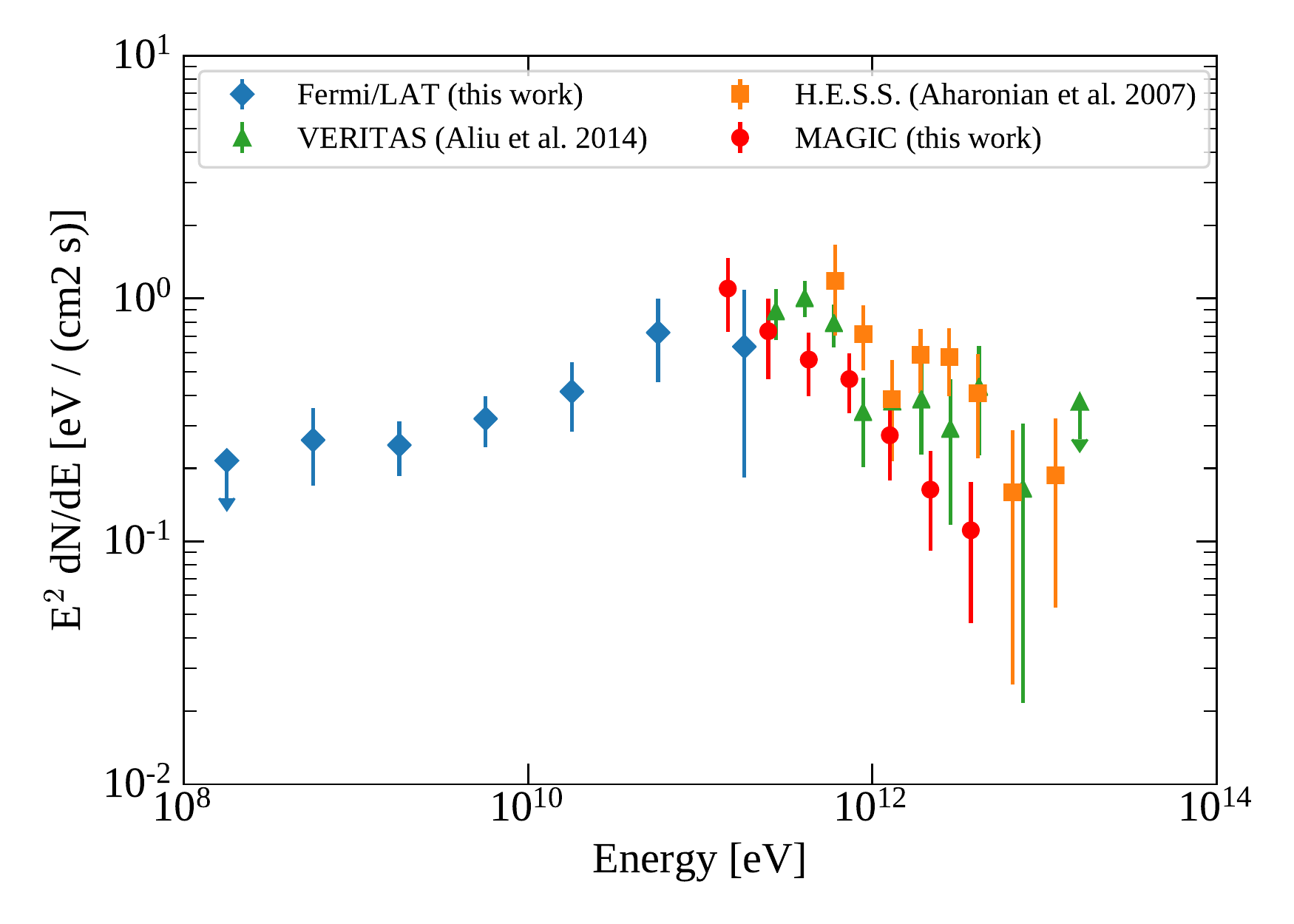}
    \caption{Spectral energy distribution of \es\ in the 100~MeV -- 100~TeV energy range. \lat\  and MAGIC data were obtained here; H.E.S.S and VERITAS measurements are taken from \citet{HESS_1ES0229} and \citet{VERITAS_1ES0229} correspondingly.}
    \label{fig:spectrum}
\end{figure}

\begin{figure}
    \includegraphics[width=\linewidth]{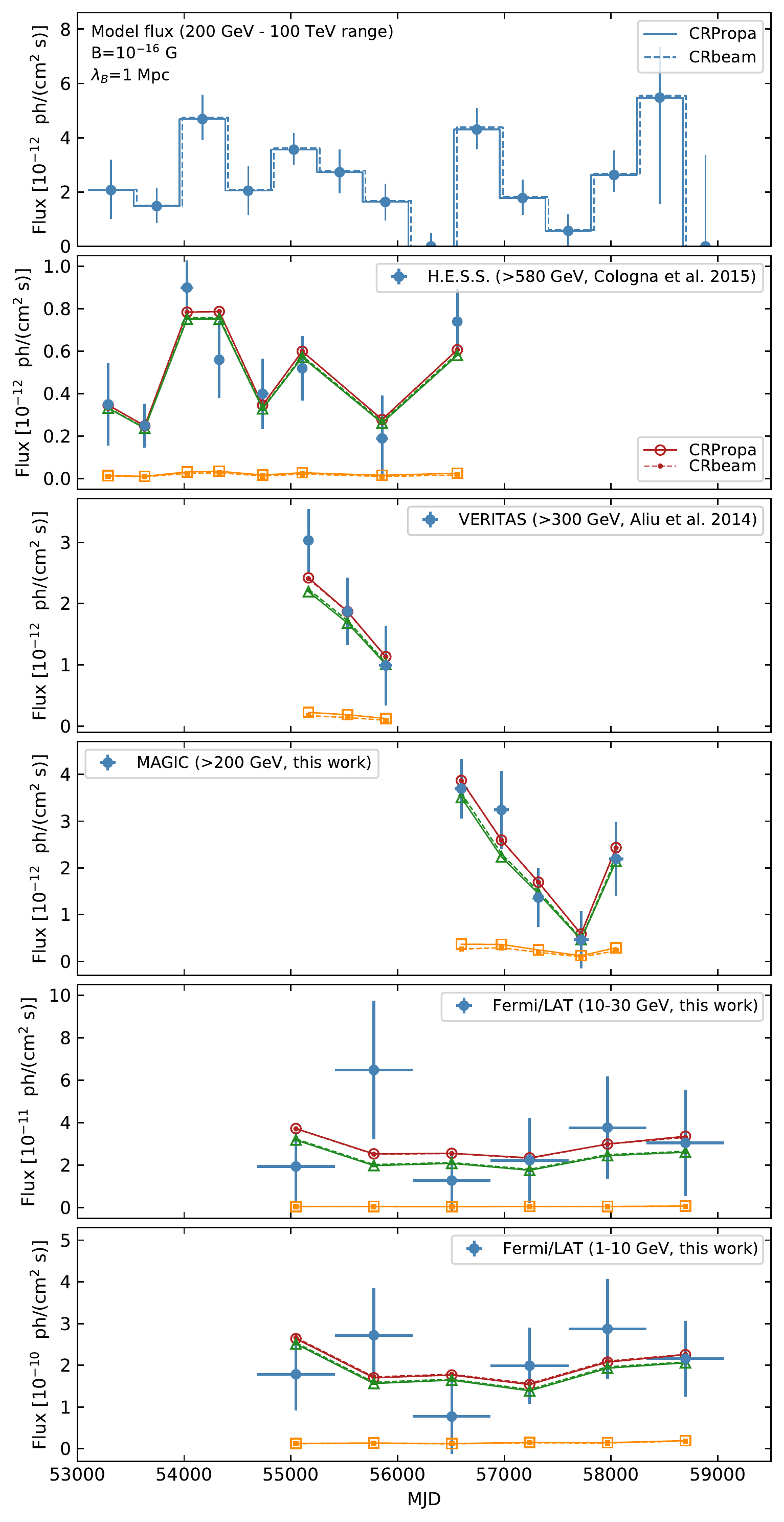}
    \caption{Light curve of \es\ in several energy bands along with an exemplary fit with IGMF of strength of $\mathrm{B=10^{-16}}$~G and coherence scale of $\mathrm{\lambda_B = 1}$~Mpc. Top panel represents the best-fit model light curve (along with its uncertainties), used to make the predictions in the energy bands where the measurements were taken (the panels below). \lat\  and MAGIC data are reported in the text; H.E.S.S. and VERITAS measurements are taken from~\citet{HESS_VERITAS_lightcurves} and~\citet{VERITAS_1ES0229} correspondingly. The primary, cascade and total source fluxes are denoted with green triangles, orange squares and red circles correspondingly. Solid and dashed lines represent calculations with CRPropa~\citep{CRPropa} and CRbeam~\citep{Berezinsky:2016feh} Monte Carlo codes respectively; the latter use the small point-like markers to distiguish themselves.}
    \label{fig:lightcurve}
\end{figure}

Spectral energy distribution and light curves of \es\ in the GeV-TeV energy range, obtained here, are shown in Fig.~\ref{fig:spectrum} and Fig.~\ref{fig:lightcurve} (see Sect.~\ref{sect:lightcurve_modelling} for full description), respectively, along with the archival measurements from H.E.S.S.~\citep{HESS_1ES0229} and VERITAS~\citep{VERITAS_1ES0229}.

The joint \lat\  and MAGIC spectrum is well described by a simple EBL-absorbed power law with cut-off model  ($\mathrm{\chi^2=2.5 / 8~\mathrm{d.o.f.}}$) with best-fit intrinsic values of $\mathrm{\Gamma = 1.72 \pm 0.05}$ and $\mathrm{E_{cut} > 2.6}$~TeV at 95\% confidence level.
In general, MAGIC, H.E.S.S. and VERITAS spectra in the $\mathrm{0.1-10}$~TeV energy range agree well with each other in shape, though differ in normalisation. The joint spectrum (including \lat\  measurements) is still described well with the same simple spectral model  ($\mathrm{\chi^2=14.2 / 22~\mathrm{d.o.f.}}$) with a compatible spectral index $\mathrm{\Gamma = 1.74 \pm 0.07}$ and a larger cut-off energy  ($\mathrm{E_{cut} > 10}$~TeV at 95\% confidence level).

The source's light curve below $\mathrm{\sim 100}$~GeV from \lat\  data does not indicate any significant time variability of the flux. On the contrary, the TeV band measurements from H.E.S.S., VERITAS and MAGIC suggest a notable variations with respect to the mean flux by a factor of $\mathrm{\sim 2}$. The flux is variable on the time scales of $\mathrm{\sim 500}$~days with several brightening and dimming episodes identifiable. The significance of this variability was assessed from a joint $\mathrm{\chi^2}$ fit of H.E.S.S., VERITAS and MAGIC light curves, using an EBL-absorbed power law with cut-off spectral model to compute the fluxes in the respective energy bins. The parameters of the model were left free in the fit. In addition, a 11\% point-to-point systematical uncertainly, characteristic for MAGIC~\citep{MAGIC_performance}, and a 25\% inter-instrument calibration uncertainty (corresponding to a typical $\mathrm{\sim 15\%}$ uncertainty on the IACT energy scale and a power law source spectrum with the measured index $\mathrm{\Gamma=2.6}$ -- similar to that of \es\, after absorption on EBL) were also accounted for. This joint fit results in $\mathrm{\chi^2 = 47/11}$~d.o.f., which rejects the constant flux assumption at the $\mathrm{4.8\sigma}$ level. MAGIC data alone, with monthly binning, hint the variability at a marginal $\mathrm{2.7\sigma}$ level  ($\mathrm{\chi^2 = 14.2 / 4~\mathrm{d.o.f.}}$) when the measurement systematical uncertainty is taken into account. Estimated fractional variability and corresponding $\mathrm{\chi^2}$ contributions to the joint fit are given in Tab.~\ref{tab::variability_summary}.

\begin{table}
    \centering
    \begin{tabular}{ccccc}
        \hline
        Name & $\mathrm{\chi^2}$ & $\mathrm{F_{var}}$ & $\mathrm{N_{points}}$ & Scale \\
        \hline
        H.E.S.S. & 22.0 & $\mathrm{0.31 \pm 0.04}$ & 8 & $\mathrm{0.85 \pm 0.23}$ \\
        MAGIC & 16.7 & $\mathrm{0.43 \pm 0.07}$ & 5 & $\mathrm{1.00}$ (fixed) \\
        VERITAS & 8.6 & $\mathrm{0.35 \pm 0.11}$ & 3 & $\mathrm{1.27 \pm 0.18}$ \\
        \hline
    \end{tabular}
    \caption{Summary of the variability study with IACT data, estimated from their joint fit with the exponentially cut off power law model. $\mathrm{F_{var}}$ denotes the fractional variability~\citep{Fvar_definition}, $\mathrm{N_{points}}$ is the corresponding light curve points counts, ``Scale'' represents the scaling parameter applied to the flux values in order to account for the possible inter-instrument systematics (arbitrarily fixed to unity for MAGIC data). Best-fit spectral parameters were $\mathrm{\log_{10}A=-22.51 \pm 0.24~\mathrm{dex(1/(eV~cm^2~s))}}$, $\mathrm{\Gamma = 1.25 \pm 0.29}$, $\mathrm{E_0=100~\mathrm{GeV}}$, $\mathrm{\log_{10}E_{cut}=13.99 \pm 1.50~\mathrm{dex(eV)}}$. Note that these parameters were determined without an account for the measured source SED shape (using only flux measurements in the corresponding energy bins) and thus differ from those using in the cascade emission modelling performed further.
    See Sect.~\ref{sect::analysis_spectrum_lightcurve} for details.}
    \label{tab::variability_summary}
\end{table}


%

\subsection{Minimal expected cascade estimate}
\label{sect:minimal_cascade}

%

Following the arguments of \citet{Neronov10}, a conservative lower bound on the IGMF strength should be based on the minimal  possible cascade contribution allowed by the data -- e.g. the softest intrinsic spectrum with the lowest cut-off
\begin{equation}
\frac{dN}{dE} = A \left(\frac{E}{E_0}\right)^{-\Gamma_{low}} \exp\left(-\frac{E}{E_{cut}^{low}}\right).
\end{equation}
In order to estimate the values of $\mathrm{\Gamma_{low}}$ and $\mathrm{E_{cut}^{low}}$ we fitted the GeV-TeV spectrum of the source, scanning the $\mathrm{\Gamma-E_{cut}}$ space by means of $\mathrm{\chi^2}$. In the scan we have combined the GeV data from \lat\  and TeV measurements of MAGIC, H.E.S.S. and VERITAS. Each spectrum was allowed to have a different normalisation during the fit, accounting for variability of the source. For each combination of $\mathrm{\Gamma-E_{cut}}$ we also computed the expected cascade power in terms of the total absorbed flux. This allows us to select the intrinsic spectrum parameters that lead to the minimal possible cascade contribution in the $\mathrm{\lesssim 100}$~GeV energy band. 

\begin{figure}
\includegraphics[width=\linewidth]{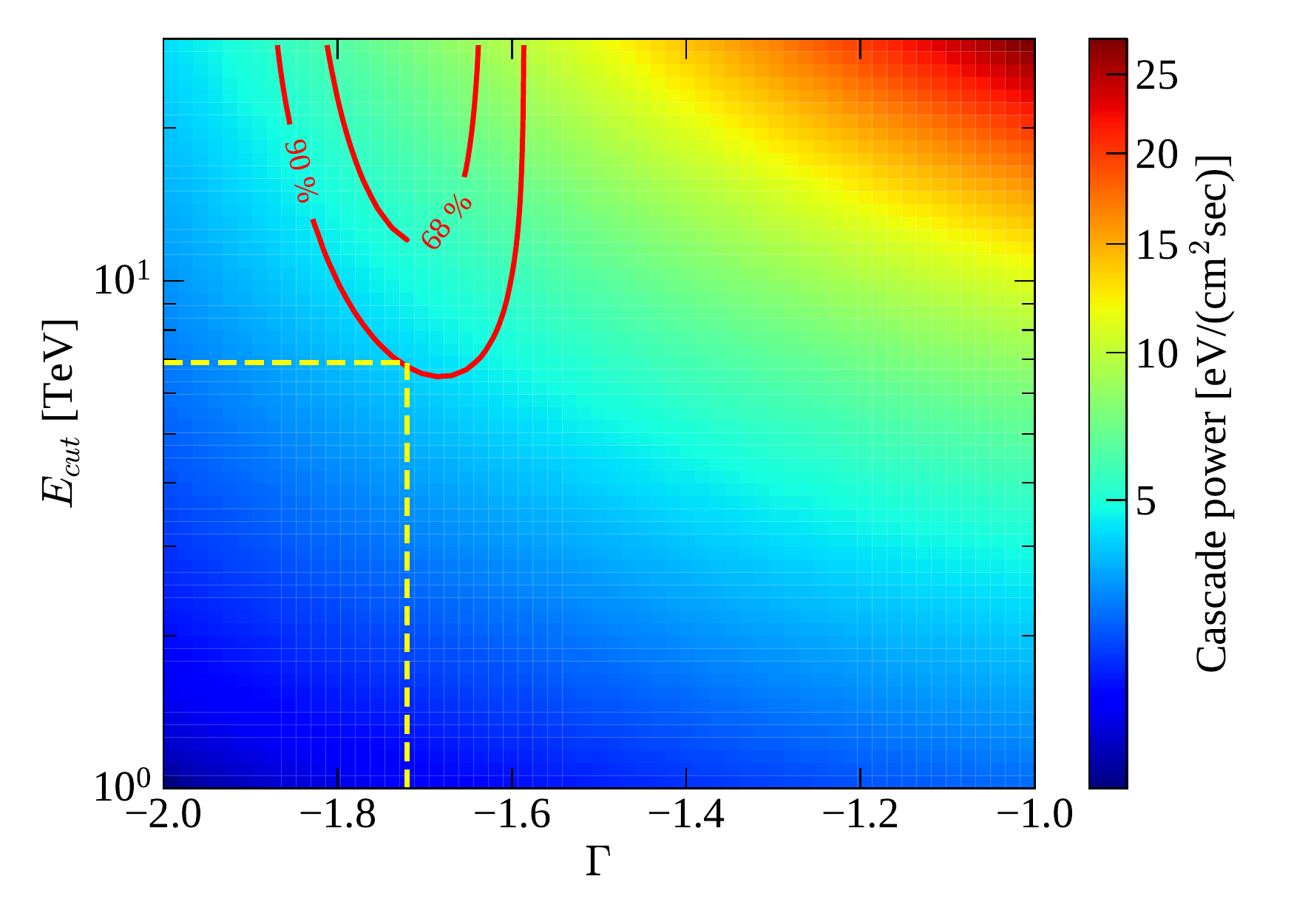}
\caption{The scan of the cascade power in the $\mathrm{\Gamma-E_{cut}}$ parameter space along with the 68\% and 90\% confidence contours from the $\mathrm{\chi^2}$ fit. At 90\% confidence level the minimal cascade, marked with the yellow dashed lines, corresponds to $\mathrm{\Gamma \approx 1.72}$ and $\mathrm{E_{cut} \approx 6.9}$~TeV.}
\label{fig:Ecut_scan}
\end{figure}

The outcome of this scan is shown in Fig.~\ref{fig:Ecut_scan}. The best $\mathrm{\chi^2}$ values correspond to $\mathrm{E_{cut} \gg 10}$~TeV and a hard spectrum with $\mathrm{\Gamma \approx -1.70}$. Still, at a 90\% confidence level the lowest cascade flux is provided by the intrinsic source spectrum with $\mathrm{\Gamma_{low} \approx -1.72}$ and $\mathrm{E_{cut}^{low} \approx 6.9}$~TeV (in the observer reference frame, otherwise it should be corrected for the source redshift by a factor of $\mathrm{[1+z]}$).

\subsection{Light curve modelling with account of IGMF}
\label{sect:lightcurve_modelling}

To model the GeV-TeV light curves, we used the model introduced in Sect.~\ref{sect:numeric_modelling}. For each IGMF configuration (i.e. strength and coherence length) we performed a fit of all H.E.S.S., VERITAS, MAGIC and \lat\  light curves together. This combination of TeV and GeV data allowed to self-consistently estimate the ``primary'' and ``delayed'' flux contribution in each time and energy bin.

In our "minimal cascade" modelling we assumed zero intrinsic source flux before $\mathrm{\mathrm{MJD} = 53000}$ (just before the first H.E.S.S. flux data point in Fig.~\ref{fig:lightcurve}). Non-zero flux at $\mathrm{\mathrm{MJD}~<~53000}$ would produce extra cascade flux during the observation time span $\mathrm{\mathrm{MJD}~>~53000}$. In addition, we have used the $\mathrm{\Gamma}$ and $\mathrm{E_{cut}}$ values that minimize the expected delayed contribution, estimated in Sect.~\ref{sect:minimal_cascade}. The constructed model is the most conservative with respect to the IGMF constraints.

The "minimal cascade" model is not free from uncertainties e.g. on the source jet orientation and opening angle. The characteristic delay of the off-centre emission of \es\ is $\mathrm{T_{d} \simeq 1 \left( \theta / 10^{-3}~\mathrm{deg} \right)^2~\mathrm{yr}}$ (with $\mathrm{\theta}$ being the offset angle); this value is set by the geometry of the $\mathrm{e^+ / e^-}$ pairs deflection and is independent from the IGMF configuration. For the time delays allowed to be probed by the data at hand -- below $\mathrm{\approx 16}$~years -- the characteristic off-centre angle of the delayed secondary emission would be $\mathrm{\lesssim 4 \times 10^{-3}}$~deg -- much less than PSF of the considered instruments or a typical opening angle of AGN jets.

\begin{figure}
    \includegraphics[width=\linewidth]{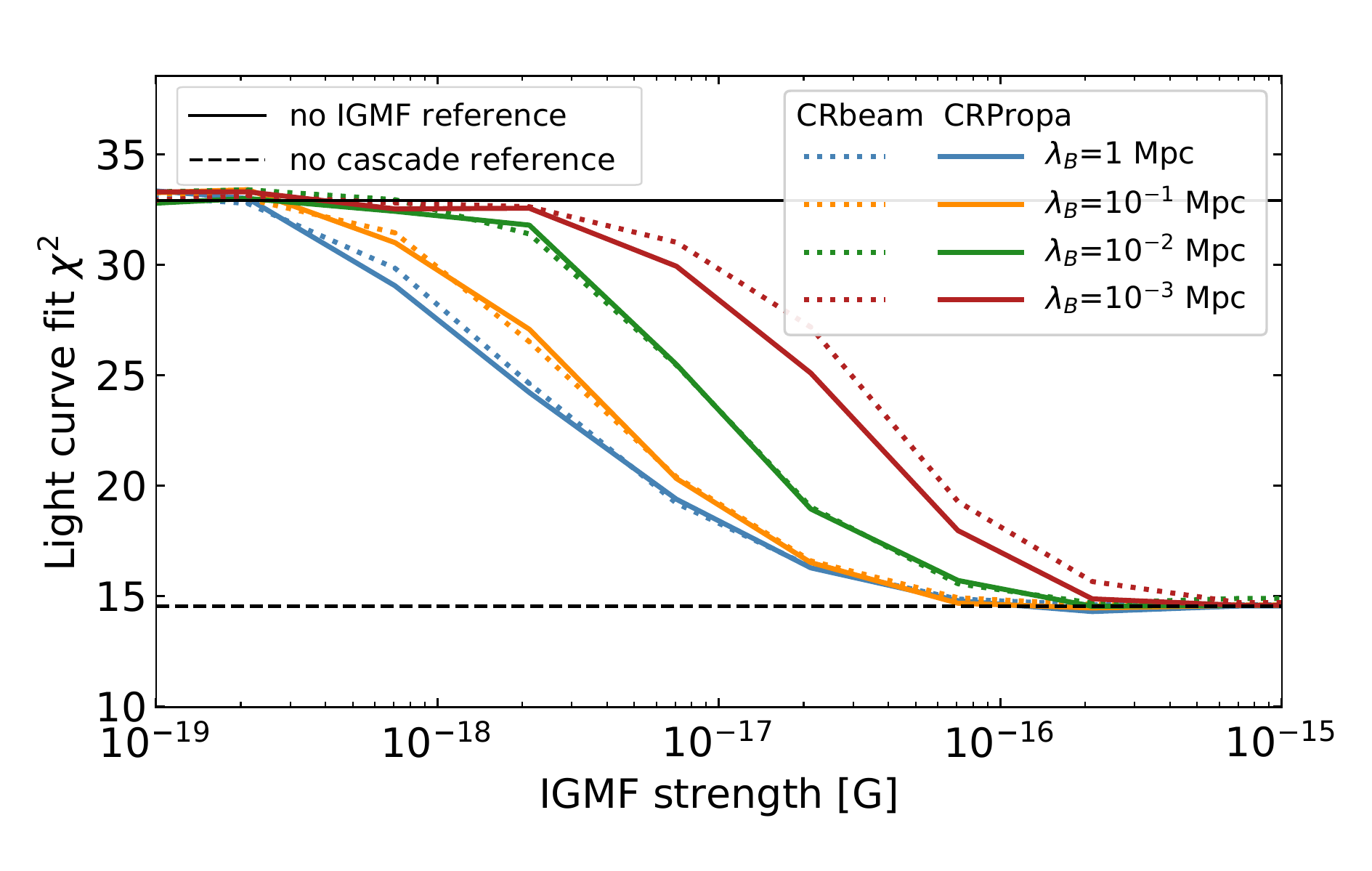}
    \caption{IGMF strength scan for several different assumed coherence lengths $\mathrm{\lambda_B}$. The scan is performed via a simultaneous fit of GeV-TeV observations at hand. Calculations with both CRPropa~\citep{CRPropa} and CRbeam~\citep{Berezinsky:2016feh} Monte Carlo codes are presented. Lines of different colors depict the $\mathrm{\chi^2}$ values estimated for different non-zero IGMF strengths. Black solid line represents the case of zero IGMF, in case of which secondary emission dominates in the \lat\  energy range; black dashed line represent the reference fit with the cascade contribution disabled.}
    \label{fig:igmf_scan}
\end{figure}

The result of the IGMF strength scan, assuming a randomly oriented IGMF with a coherence scale in the range $\mathrm{\lambda_B=10^{-3} - 1}$~Mpc, is shown in Fig.~\ref{fig:igmf_scan}. For IGMF strengths below $\mathrm{\sim 10^{-18}}$~G the secondary emission suffers minimal lag (below $\mathrm{\sim 1}$~yr), so that there is almost no suppression due to the time delay. For IGMF stronger than $\mathrm{\sim 10^{-16}}$~G the time delay is large enough to suppress the secondary emission well below the expected primary one, so that the model light curves at GeV energies start to simply mirror the variability at higher-energy, TeV spectrum end. The minimal $\mathrm{\chi^2}$ value here represents the ability of the constructed model to reproduce the short time scale  ($\mathrm{< 400}$~days) variability of \es, suggested by the data points in Fig.~\ref{fig:lightcurve}.

This scan in Fig.~\ref{fig:igmf_scan} indicates that with very few assumptions made here, the IGMF is constrained to have strength $\mathrm{B \gtrsim 1.8\times 10^{-17}}$~G at 95\% confidence level (corresponding to $\mathrm{\Delta \chi^2 \approx 2.71}$ from the the minimal $\mathrm{\chi^2}$ value).

The scaling of the above constraint with the IGMF correlation length is set by the cooling distance of the injected electrons and positrons and for randomly oriented IGMF cells can be derived analytically~\citep{NeronovSemikoz09}. The cooling distance of $\mathrm{e^+ / e^-}$ pairs (due to radiative loss through inverse Compton scattering) is
\begin{equation}
  l_{IC}\simeq 160\left[\frac{E_{IC}}{3\mbox{ GeV}}\right]^{-1/2}\mbox{ kpc}
\end{equation}
for electrons and positrons up-scattering CMB photons to the energy $\mathrm{E_{IC}}$. The scan performed above is most sensitive to the delayed emission contribution in the lowest energy 1-10~GeV light curve. This suggests that the derived lower limit holds for IGMF coherence scales $\mathrm{\lambda_B > 0.2~\mathrm{Mpc}}$ and scales for shorter correlation lengths as $\mathrm{\lambda_B^{1/2}}$, so that it can be summarised as:
\begin{equation}
    B \gtrsim \left\{
    \begin{array}{lr}
        1.8 \times 10^{-17}~\mathrm{G} &, \lambda_B > 0.2~\mathrm{Mpc} \\
        1.8 \times 10^{-17} \left( \lambda_B / 0.2~\mathrm{Mpc} \right)^{-1/2}~\mathrm{G} &, \lambda_B < 0.2~\mathrm{Mpc} \\
    \end{array}
    \right.
    \label{eq::igmf_limit}
\end{equation}
This conclusion is supported by the scan of the simulated $\mathrm{\lambda_B~<~1}$~Mpc cases that we performed.

Worthy to note that the presented analysis does not favour a specific value of $\mathrm{B}$ or $\mathrm{\lambda_B}$ as every IGMF satisfying the limit in Eq.~\ref{eq::igmf_limit} would result in the same secondary flux time delay. This degeneracy, in principle, can be broken if an accurate measurement of the cascade flux time evolution is obtained~\citep{Neronov13}; in the absense of such a firm detection, we are not in a position to do this here.

\subsection{Effect of the possible spectral variations}

The assumption of the fixed source spectral shape, used here, may not be justified if \es\, spectral shape varies over the 14 years of \lat\, observations. However, no indications for its change have been found in earlier H.E.S.S. and VERITAS data with possible spectral index variations $\mathrm{\Delta \Gamma \lesssim 0.2}$~\citep{HESS_1ES0229, VERITAS_1ES0229}; we have not found such indications in the MAGIC data either. Such small variations do not have a noticeable impact on the derived IGMF bound value.

Still, given the lower source flux during the MAGIC observations~(see Fig.~\ref{fig:lightcurve}), the cut-off energy is poorly constrained for time period MJD~$56500-58000$ and may be substantially lower than the assumed $\mathrm{E_{cut}^{low} \approx 6.9}$~TeV. Conservatively assuming that no detectable cascade was thus generated during the MAGIC observations, we have found that lower limit on IGMF strength is relaxed to $\mathrm{B\gtrsim 6\times 10^{-18}}$~G for $\mathrm{\lambda_B > 0.2~Mpc}$.


\section{Discussion and conclusions}
\label{sect::discussion}

A range of lower bounds on the IGMF strength has been previously reported based on non-observation of delayed emission in the 1-100 GeV band, assuming that the TeV sources remain active on year-to-decade time scales \citep{dermer11,Taylor11,Vovk12,fermi_limit}\footnote{Note that the difference between the lower bounds derived by \citet{dermer11} and \citet{Taylor11}, $\mathrm{10^{-18}}$~G and $\mathrm{10^{-17}}$~G is due to more precise modelling of the cascade emission by \citet{Taylor11} (Monte Carlo) compared to  \citet{dermer11}  (approximate analytic model).}. An important limitation of all previously reported bounds is that none of the previously reported analyses used strictly contemporaneous monitoring of the source in the 1-100 GeV and TeV bands.

The bound derived in the present paper is based on a combination of long-term simultaneous monitoring  of \es\ in the 1-100 GeV energy range with \lat\  and in the $\mathrm{E>200}$~GeV range with Cherenkov telescopes: H.E.S.S., VERITAS and MAGIC. This combination alleviates for the first time the dependence of the bound on an uncertainty related to possible variability of the TeV band source flux. This makes the bound more robust and almost free from uncertainties of the intrinsic primary source flux variability.

We would like to note that a certain fraction of the deposited electron-positron power may be carried away via plasma instabilities, developing as a result of interaction between the generated particle stream and the intergalactic medium~\citep{broderick12}. Such instabilities may mimic the effect of IGMF, reducing the expected cascade flux. Presently there is no self-consistent description of the problem given the large density gradient between the stream and surrounding plasma, so that conclusions on its importance for the IGMF measurements vary substantially depending on the assumptions adopted~\citep{broderick12, miniati13, chang14, shalaby18, vafin18, vafin19, batista19, perry21}. Furthermore, the instability development may be affected by the source emission lifetime~\citep{broderick12} and by IGMF itself~\citep{alawashra22}. Given these uncertainties, a quantitative analysis of this issue is beyond the scope of this work.

\begin{figure}
\includegraphics[width=\linewidth]{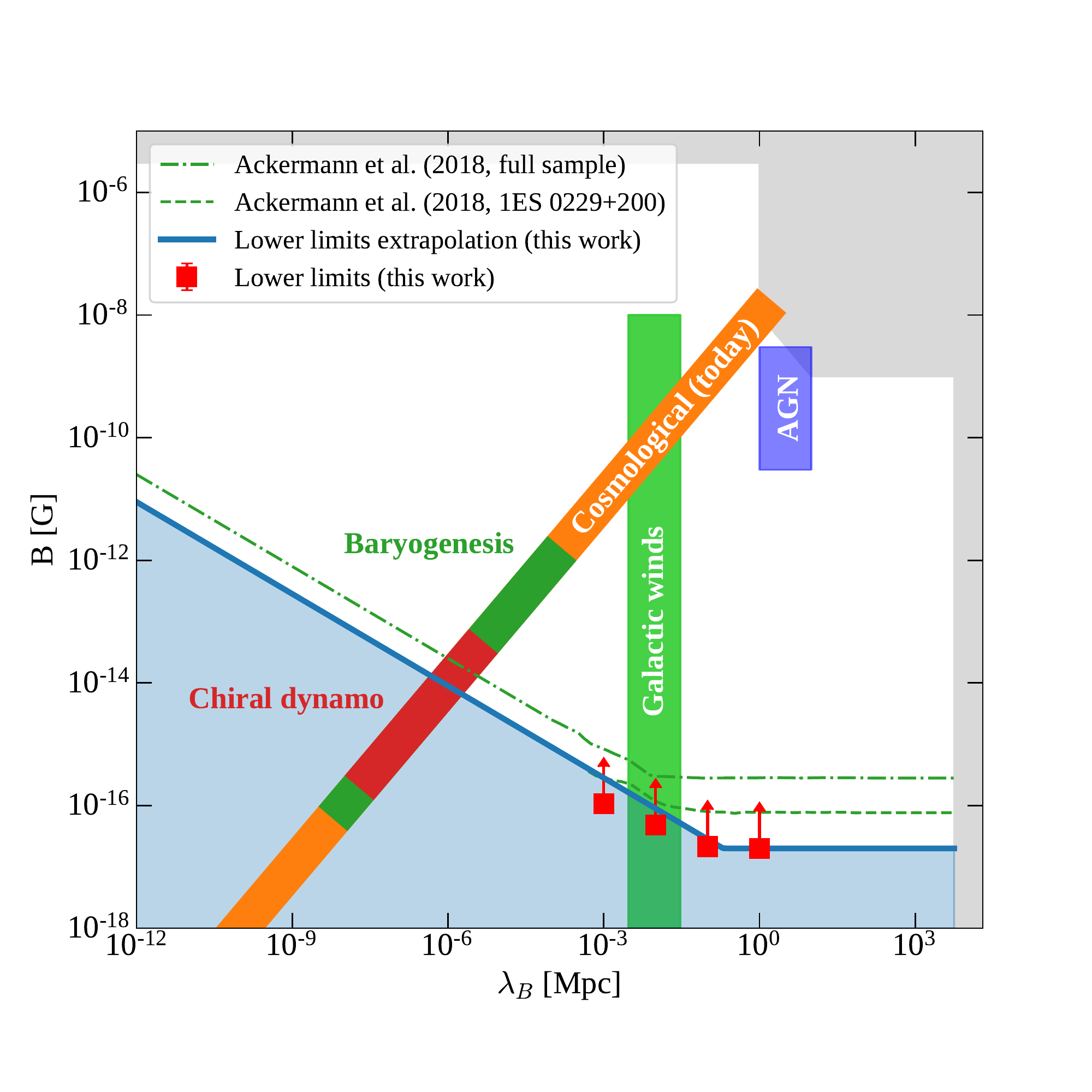}
\caption{Lower bound on IGMF strength derived from \lat\  and Cherenkov telescope data sets (thick blue curve and red data points). Green dot-dashed and dashed curves show previous \lat\  limits derived for the full source sample and \es\ only~\citep{fermi_limit}. Light-grey shaded upper bound shows previously known limits on the IGMF strength and correlation length from radio telescope data  \citep{kronberg} and CMB~\citep{planck} analysis as well as from theoretical estimates~\citep{Durrer13}. Inclined orange stripe shows the locus of end points of evolution of cosmological magnetic fields~\citep{banerjee}. Red stripe marks possible range of magnetic field produced by the chiral dynamo~\citep{joyce,neronov20}. Dark green stripe denotes the range of electroweak phase transition magnetic fields which might explain the observed baryonic asymmetry of the Universe ~\citep{shaposhnikov98,fujita16,kamada16}. Filled vertical green and violet boxes show favored regions of IGMF generated by a frozen-in magnetic field, originating from AGN outflows~\citep{Furlanetto:2001} or galactic winds~\citep{Bertone:2006} as labelled in the figure.}
\label{fig:exclusion}
\end{figure}

Fig.~\ref{fig:exclusion} puts our bound in the context of other measurements and theoretical models of IGMF. The cosmological evolution of magnetic fields which might have been present in the Early Universe drives the field strength and correlation length toward 
\begin{equation}
    B_{\mathrm{cosmological}}\simeq 10^{-8}\left(\frac{\lambda_B}{1\mbox{ Mpc}}\right)\mbox{ G}
    \label{eq::cosmological_igmf}
\end{equation}
shown by the inclined orange band in Fig.~\ref{fig:exclusion} \citep{banerjee, Durrer13}. Non-observation of Faraday rotation of the radio waves polarization from distant active galactic nuclei and of magnetic field imprint on the cosmic microwave background constrains the IGMF strength from above for large $\mathrm{\lambda_B}$ \citep{kronberg, Durrer13}. \gr\ observations reported here constrain the field from below. For the particular case of cosmological magnetic field, equating Eqs.~\ref{eq::igmf_limit} and~\ref{eq::cosmological_igmf} we find
\begin{equation}
    B_{\mathrm{cosmological}} \gtrsim 10^{-14}\mbox{ G.}
\end{equation} 

The bound derived here (Eq.~\ref{eq::igmf_limit}) is weaker than that reported by \citet{fermi_limit} based on the stacking analysis of a number of blazars. However one should note that the effect of the IACT and \lat\  energy scale calibration uncertainties relaxes that bound threefold~\citep{fermi_limit}. Moreover, the IGMF limit presented here is based on a prior conservative estimate of the intrinsic \es\ spectrum, aimed at minimizing the expected cascade regardless of IGMF. Finally, the bound of  \citet{fermi_limit} was relying on a strong assumption about properties of the TeV band \gr\ fluxes of the stacked sources, rather than on real measurements of the TeV flux. In contrast, the bound reported here is free from this assumption. Instead, it relies on precise measurements of the flux history of the source in the GeV-TeV energy range.

The robust lower bound on IGMF strongly constrains a range of testable models of cosmological magnetogenesis -- and specifically the models of magnetic field production at / before the electroweak phase transition (EWPT) in the hot Universe at the temperature close to $\mathrm{T\sim 100}$~GeV, during the first nano-second after the Big Bang~\citep[see][and references therein]{NeronovSemikoz09,Durrer13}.  These models are particularly interesting in the context of the problem of the origin of baryon asymmetry of the Universe (BAU). It is possible that the BAU has been generated through the transfer of hypermagnetic helicity to the baryon number at the moment of EWPT~\citep{shaposhnikov98,fujita16,kamada16}. If this is the case, the hypermagnetic field should have been present in the Universe before the EWPT, when the temperature was still higher. This is possible, for example, through the ``chiral dynamo'' magnification of thermal fluctuations of hypermagnetic field at temperatures $\mathrm{T>80}$~TeV \citep{joyce,neronov20}. Estimates of possible initial parameters of the cosmological hypermagnetic field at the moment of chiral dynamo action are shown by the red stripe in Fig.~\ref{fig:exclusion}. At the moment of the EWPT the chiral dynamo field parameters fall into a range of fields needed for successful BAU generation~\citep{fujita16,kamada16}. Subsequent evolution of the field to the present epoch results in the cosmological fields with strength from $\mathrm{\sim 10^{-15}}$~G to $\mathrm{\sim 5\times 10^{-14}}$~G, close to the lower bound derived here.

The analysis presented above indicates that the chiral dynamo magnetic field, possibly responsible for BAU, can be measured using the technique of the search of time-delayed emission in the signal of extragalactic sources of TeV \GR s. In the specific case of sources at the redshift $\mathrm{z\sim 0.1}$ similar to that of \es, detection of the $\mathrm{\sim 10}$~yr delayed signal requires systematic monitoring of the sources in both the TeV and in 1-100~GeV bands.

Shorter time delays are expected at higher energies, because the delay time shortens as $\mathrm{1/E_\gamma^2}$~\citep{NeronovSemikoz09}. In the specific case of \es\ it goes down to $\mathrm{\sim 1}$~yr at $\mathrm{100}$~GeV. This opens a possibility of detection of the BAU-related magnetic field through regular monitoring of source variability at intra-year time scale with Cherenkov telescopes. Detection of a specifically energy-dependent ``afterglow'' of a bright short pronounced flare of the source would be a ``smoking gun'' of the effect in question. Unfortunately, \es\ does not exhibit intense flaring episodes. Other candidate sources could be considered for this approach. A necessary condition is that the candidate sources should have strong intrinsic luminosity above 10~TeV~\citep[the energy of primary \GR s which induce cascade emission at 100~GeV, ][]{NeronovSemikoz09}. Measurement of such time delays through long-term monitoring is challenging, but not impossible with existing and planned \gr\ observation facilities including \lat\ ~\citep{FermiLAT}, CTA~\citep{cta} and HERD~\citep{herd}.


\begin{acknowledgements}
We would like to thank the Instituto de Astrof\'{\i}sica de Canarias for the excellent working conditions at the Observatorio del Roque de los Muchachos in La Palma. The financial support of the German BMBF, MPG and HGF; the Italian INFN and INAF; the Swiss National Fund SNF; the ERDF under the Spanish Ministerio de Ciencia e Innovaci\'on (MICINN) (PID2019-104114RB-C31, PID2019-104114RB-C32, PID2019-104114RB-C33, PID2019-105510GB-C31,PID2019-107847RB-C41, PID2019-107847RB-C42, PID2019-107847RB-C44, PID2019-107988GB-C22); the Indian Department of Atomic Energy; the Japanese ICRR, the University of Tokyo, JSPS, and MEXT; the Bulgarian Ministry of Education and Science, National RI Roadmap Project DO1-400/18.12.2020 and the Academy of Finland grant nr. 320045 is gratefully acknowledged. This work was also supported by the Spanish Centro de Excelencia ``Severo Ochoa'' (SEV-2016-0588, SEV-2017-0709, CEX2019-000920-S), the Unidad de Excelencia ``Mar\'{\i}a de Maeztu'' (CEX2019-000918-M, MDM-2015-0509-18-2) and by the CERCA program of the Generalitat de Catalunya; by the Croatian Science Foundation (HrZZ) Project IP-2016-06-9782 and the University of Rijeka Project uniri-prirod-18-48; by the DFG Collaborative Research Centers SFB823/C4 and SFB876/C3; the Polish National Research Centre grant UMO-2016/22/M/ST9/00382; and by the Brazilian MCTIC, CNPq and FAPERJ.
\end{acknowledgements}

{
    \tiny
    \noindent \textit{Author contributions}
    Ie.Vovk: methodology, simulations, Fermi/LAT and MAGIC data reduction/analysis, statistical analysis;
    A.Neronov: methodology;
    P. Da Vela: MAGIC data analysis;
    A.Stamerra: project supervision, coordination of MAGIC observations, MAGIC data analysis;
    D.Semikoz: interpretation;
    A.Korochkin: simulations, statistical analysis.
    The rest of the authors have contributed in one or several of the following ways: design, construction, maintenance and operation of the MAGIC telescopes; preparation and/or evaluation of the observation proposals; data acquisition, processing, calibration and/or reduction; production of analysis tools and/or related Monte Carlo simulations; discussion and approval of the contents of the draft.

}

\bibliography{0229_MAGIC}

\end{document}